\documentclass{llncs}
\usepackage{amssymb,amsmath,amscd,latexsym,epic,eepic}
\usepackage{algorithm}
\usepackage{paralist}
\def\qed{\rule{0.4em}{1.4ex}}

\newcommand{\restr}{\upharpoonright}
\newcommand{\set}[1]{\{#1\}}
\newcommand{\sseq}{\langle s_0,s_1,s_2,\ldots\rangle}

\newcommand{\pat}{\omega}
\newcommand{\Pat}{\Omega}
\newcommand{\Paths}{\Omega}
\newcommand{\straa}{\sigma}
\newcommand{\Straa}{\Sigma}
\newcommand{\strab}{\pi}
\newcommand{\Strab}{\Pi}
\newcommand{\ct}{\mathsf{ct}}

\newcommand{\rank}{\mathit{rank}}

\newcommand{\Attr}{\mathit{Attr}}
\newcommand{\attr}{\mathit{Attr}}

\newcommand{\Occur}{\mathrm{Ocuur}}
\newcommand{\wpe}{\mathrm{WeakParityEven}}
\newcommand{\wpo}{\mathrm{WeakParityOdd}}

\newcommand{\Reach}[1]{\mathrm{Reach}(#1)}
\newcommand{\Safe}[1]{\mathrm{Safe}(#1)}

\newcommand{\nats}{\mathbb{N}}

\title{Linear Time Algorithm for Weak Parity Games 
}
\author{%
Krishnendu Chatterjee%
}
\institute{%
University of California, Berkeley, USA\\
{\tt c\_krish@eecs.berkeley.edu}
}
\date{}
\begin{document}

\maketitle
\begin{abstract}
We consider games played on graphs with 
the winning conditions for the players specified as \emph{weak-parity}
conditions.
In weak-parity conditions the winner of a play is decided by 
looking into the set of states appearing in the play, rather
than the set of states appearing infinitely often in the play.
A naive analysis of the classical algorithm for weak-parity
games yields a quadratic time algorithm.
We present a linear time algorithm for solving weak-parity games.
\end{abstract}

\section{Introduction}
We consider two-player games on graphs with winning objectives formalized as
a \emph{weak-parity} objective~\cite{Thomas97}.
In a two-player game~\cite{Mar75}, the set of vertices or states are partitioned into
player~1 states and player~2 states.
At player~1 states player~1 decides the successor and likewise for player~2.
We consider weak-parity objectives, where we have a priority function that
maps every state to an integer priority.
A \emph{play} is an infinite sequence of states, and in a weak-parity objective
the winner of a play is decided by considering the minimum priority state
that appear in the play: if the minimum priority is even, then player~1 wins,
and otherwise player~2 is the winner.
The classical algorithm to solve weak-parity games with a naive running 
time analysis works in $O(d \cdot m)$ time, where $d$ is the number of
priorities and $m$ is the number of edges of the game graph.
Since $d$ can be $O(n)$, in the worst case the naive analysis requires $O(n \cdot m)$ 
time, where $n$ is the number of states.
We present an improved analysis of the algorithm and show that the algorithm 
works in $O(m)$ time.

\section{Definitions}
We consider turn-based deterministic games played by two-players with 
\emph{weak-parity} objectives; we call them 
weak-parity games.
We define game graphs, plays, strategies, objectives and notion of winning 
below.

\medskip\noindent{\bf Game graphs.} 
A \emph{game graph} $G=((S,E),(S_1,S_2))$ consists of a directed graph 
$(S,E)$ with a finite state space $S$ and a set $E$ of edges, 
and a partition $(S_1,S_2)$ of the state space $S$ into two sets.
The states in $S_1$ are player~1 states, and the states in $S_2$ are 
player~2 states.
For a state $s\in S$, we write $E(s)=\set{t\in S \mid (s,t) \in E}$ 
for the set of successor states of~$s$.
We assume that every state has at least one out-going edge, 
i.e., $E(s)$ is non-empty for all states $s\in S$.

\medskip\noindent{\bf Plays.}
A game is played by two players: 
player~1 and player~2, who form an
infinite path in the game graph by moving a token along edges.
They start by placing the token on an initial state, and then they
take moves indefinitely in the following way.
If the token is on a state in~$S_1$, then player~1 moves the token along
one of the edges going out of the state.
If the token is on a state in~$S_2$, then player~2 does likewise.
The result is an infinite path in the game graph;
we refer to such infinite paths as plays.
Formally, a \emph{play} is an infinite sequence 
$\sseq$ of states such that $(s_k,s_{k+1}) \in E$ for all $k \geq 0$. 
We write $\Pat$ for the set of all plays.

\medskip\noindent{\bf Strategies.} 
A strategy for a player is a recipe that specifies how to extend plays.
Formally, a \emph{strategy} $\straa$ for player~1 is a function 
$\straa$: $S^* \cdot S_1 \to S$ that, given a finite sequence of states 
(representing the history of the play so far) which ends in a player~1 
state, chooses the next state.
The strategy must choose only available successors, i.e., for all $w \in S^*$ 
and $s \in S_1$ we have $\straa(w \cdot s) \in E(s)$.
The strategies for player~2 are defined analogously.
We write $\Straa$ and $\Strab$ for the sets of all strategies for 
player~1 and player~2, respectively.
An important special class of strategies are \emph{memoryless} strategies.
The memoryless strategies do not depend on the history of a play, 
but only on the current state. 
Each memoryless strategy for player~1 can be specified as a function
$\straa$: $S_1 \to S$ such that $\straa(s) \in E(s)$ for all $s \in S_1$,
and analogously for memoryless player~2 strategies.
Given a starting state $s\in S$, a strategy $\straa\in\Straa$ for player~1, 
and a strategy $\strab\in\Strab$ for player~2, there is a unique play, 
denoted $\pat(s,\straa,\strab)=\sseq$, which is defined as follows: 
$s_0=s$ and for all $k \geq 0$,
if $s_k \in S_1$, then $\straa(s_0,s_1,\ldots,s_k)=s_{k+1}$, and
if $s_k \in S_2$, then $\strab(s_0,s_1,\ldots,s_k)=s_{k+1}$.

\medskip\noindent{\bf Weak-parity objectives.}
We consider game graphs with weak-parity objectives for player~1 
and the complementary weak-parity objectives for player~2.
For a play $\pat = \sseq\in \Omega$,   
we define $\Occur(\pat) = 
\set{s \in S \mid \mbox{$s_k = s$ for some $k \geq 0$}}$
to be the set of states that occur in~$\pat$.
We also define reachability and safety objectives as they will be useful in the
analysis of the algorithms.
\begin{enumerate}

 \item 
  \emph{Reachability and safety objectives.}
  Given a set $T \subseteq S$ of states, the reachability objective 
  $\Reach{T}$ requires that some state in $T$ be visited,
  and dually, 
  the safety objective $\Safe{F}$ requires that only states in $F$ 
  be visited.
  Formally, the sets of winning plays are
  $\Reach{T}= \set{\sseq \in \Paths \mid 
  \exists k \geq 0. \ s_k \in T}$
  and 
  $\Safe{F}=\set{\sseq \in \Paths \mid 
  \forall k \geq 0.\ s_k \in F}$.
  The reachability and safety objectives are dual in the sense that 
  $\Reach{T}= \Pat \setminus \Safe{S \setminus T}$.

\item \emph{Weak-parity objectives.}
  For $d \in \nats$, we let $[d] = \set{0, 1, \ldots, d-1}$ and 
  $[d]_+=\set{1,2,\ldots,d}$.
  Let $p : S \to [d]$ be a function that assigns a \emph{priority}
  $p(s)$ to every state $s \in S$. 
  The weak-parity objective requires that the minimal priority 
  occurring is \emph{even}.
  Formally, the set of winning plays is 
  $\wpe(p)=\set{\pat \in \Paths \mid \min(p(\Occur(\pat))) \text{ is even}}$.
  The complementary objective to $\wpe(p)$ is $\wpo(p)$ defined as the set 
  $\wpo(p)=\set{\pat \in \Paths \mid \min(p(\Occur(\pat))) \text{ is odd}}$ of
  winning plays.

\end{enumerate}

\medskip\noindent{\bf Winning strategies and sets.}
Given a game graph $G$ and 
an objective $\Phi\subseteq\Pat$ for player~1, a strategy 
$\straa\in\Straa$ is a \emph{winning strategy}
for player~1 from a state $s$ if for all player~2 strategies $\strab\in\Strab$ 
the play $\pat(s,\straa,\strab)$ is winning, i.e., 
$\pat(s,\straa,\strab) \in \Phi$.
The winning strategies for player~2 are defined analogously.
A state $s\in S$ is winning for player~1 with respect to the objective 
$\Phi$ if player~1 has a winning strategy from $s$.
Formally, the set of \emph{winning states} for player~1 with respect to 
the objective $\Phi$ in a game graph $G$ is 
$W_1^G(\Phi) =\set{s \in S \mid \exists \straa\in\Straa. 
\ \forall \strab\in\Strab.\ \pat(s,\straa,\strab) \in \Phi}.$
Analogously, the set of winning states for player~2 with respect to an 
objective $\Psi\subseteq\Pat$ is 
$W_2^G(\Psi) =\set{s \in S \mid \exists \strab\in\Strab. \ 
\forall \straa\in\Straa.\ \pat(s,\straa,\strab) \in \Psi}.$
If the game graph is clear from the context we drop the 
game graph from the superscript.
We say that there exists a memoryless winning strategy 
for player~1 with respect to the objective $\Phi$ if there exists such 
a strategy from all states in $W_1(\Phi)$; and similarly for player~2.

\begin{theorem}\label{thrm:determinacy}
For all game graphs $G=((S,E),(S_1,S_2))$, for all weak-parity 
objectives
$\Phi=\wpe(p)$ for player~1, and the complementary objective 
$\Psi=\Pat \setminus \Phi$ for player~2,
the following assertions hold.
\begin{enumerate}
\item  We have $W_1(\Phi)=S \setminus W_2(\Psi)$.
\item  There exists a memoryless winning strategy for both players.
\end{enumerate}
\end{theorem}

\medskip\noindent{\bf Closed sets and attractors.}
Some notions that will play key roles in the analysis of the algorithms are 
the notion of \emph{closed sets} and \emph{attractors}.
We define them below.

\smallskip\noindent{\em Closed sets.} 
A set $U\subseteq S$ of states is a \emph{closed set} 
for player~1 if the following two conditions hold:
(a)~for all states $u \in (U \cap S_1)$, we have $E(u) \subseteq U$, 
i.e., all successors of player~1 states in $U$ are again in $U$; and
(b)~for all $u \in (U \cap S_2)$, we have 
$E(u) \cap U \neq \emptyset$, i.e., 
every player~2 state in $U$ has a successor in $U$.
A player~1 closed set is also called a \emph{trap} for player~1.
The closed sets for player~2 are defined analogously.
Every closed set $U$ for player~$\ell$, for $\ell\in\set{1,2}$, 
induces a sub-game graph, denoted $G \restr U$.

\begin{proposition}\label{prop:closed}
Consider a game graph $G$, and a closed set $U$ for player~2.
For every objective $\Phi$ for player~1, 
we have $W_1^{G \restr U}(\Phi)  \subseteq W_1^G(\Phi)$.  
\end{proposition}

\smallskip\noindent{\em Attractors.} 
Given a game graph $G$, a set $U \subseteq S$ of states, 
and a player $\ell\in\set{1,2}$, 
the set $\Attr_\ell(U,G)$ contains the states from which player~$\ell$ 
has a strategy to reach a state in $U$ against all strategies
of the other player;
that is, $\Attr_\ell(U,G) = W_\ell(\Reach{U})$.
The set $\Attr_1(U,G)$ can be computed inductively as follows:
let $R_0=U$; let 
\[
R_{i+1}= R_i \cup \set{s \in S_1 \mid E(s) \cap R_i \neq \emptyset} 
    \cup \set{s \in S_2 \mid E(s) \subseteq R_i}
\qquad \text{for all } i\ge 0;
\]
then $\Attr_1(U,G)= \bigcup_{i\ge 0} R_i$.
The inductive computation of $\Attr_2(U,G)$ is analogous.
For all states $s \in \Attr_1(U,G)$, define 
$\rank(s)=i$ if $s \in R_i \setminus R_{i-1}$,
that is, $\rank(s)$ denotes the least $i\ge 0$ such that $s$ is 
included in $R_i$.
Define a memoryless attractor strategy $\straa\in\Straa$ 
for player~1 as follows: 
for each state $s \in (\Attr_1(U,G) \cap S_1)$ with $\rank(s)=i$, 
choose a successor $\straa(s)\in (R_{i-1} \cap E(s))$ 
(such a successor exists by the inductive definition).
It follows that for all states $s \in \Attr_1(U,G)$ and all strategies 
$\strab\in\Strab$ for player~2, the play $\pat(s,\straa,\strab)$ reaches 
$U$ in at most $|\Attr_1(U,G)|$ transitions.

\begin{proposition}\label{prop:attractor}
For all game graphs $G$, all players $\ell\in\set{1,2}$, and 
all sets $U\subseteq S$ of states, 
the set $S\setminus \Attr_\ell(U,G)$ is a closed set for player~$\ell$.
\end{proposition}

\medskip\noindent{\bf Notation.} For a game graph $G=((S,E),(S_1,S_2))$, 
a set $U \subseteq S$ and $\ell\in \set{1,2}$, 
we write $G\setminus \Attr_{\ell}(U,G)$ to denote the game graph 
$G \restr (S \setminus \Attr_{\ell}(U,G))$.

\medskip\noindent{\bf Computation of attractors.}
Given a game graph $G=(S,E)$ and a set $T \subseteq S$ of states let us denote 
by $A=\attr_{\ell}(T,G)$ the attractor for a player $\ell\in \set{1,2}$ 
to the set $T$. 
A naive analysis of the computation of attractor shows that the 
computation can be done in $O(m)$ time, where 
$m$ is the number of edges.
An improved analysis can be done as follows.
For every state $s \in S \setminus T$ we keep a counter initialized to~0.
Whenever a state $t$ is included for the set of states in $A$, for all states
$s$ such that $(s,t) \in E$ we increase the counter by~1.
For a state $s \in S_{\ell}$ if the counter is positive, then we include it in $A$,
and for a state $s \in S \setminus S_{\ell}$ if the counter equals
the number of edges $|E(s)|$, then we include it in $A$.
Let us consider the following set of edges: $E_A= E \cap ( (S \setminus T) \times A)$.
The work of the attractor computation is only on edges with the start state
in $(S \setminus T)$ and end state in $A$. That is the total work of attractor
computation on edges is $O(m_A)$ where $m_A=|E_A|$.
Also the counter initialization phase does not require to initialize counters for
all states,
but only initializes a counter for a state $s$, when some state $t \in E(s)$
gets included in $A$ for the first time.
This gives us the following lemma.

\begin{lemma}\label{lemm:attrcompute}
Given a game graph $G=(S,E)$ and a set $T \subseteq S$ of states let us denote 
by $A=\attr_{\ell}(T,G)$ the attractor for a player $\ell\in \set{1,2}$ 
to the set $T$.
The set $A$ can be computed in time $O(|E_A|)$ , where 
$E_A= E \cap ( (S\setminus T) \times A)$.
\end{lemma}

\section{The Classical Algorithm}
We first present the classical algorithm 
for weak-parity games and present an improved analysis
to show that the algorithm has a linear-time complexity.
We first present an informal description of the 
algorithm; and a formal description of the algorithm 
is given as Algorithm~\ref{algorithm:classical}.

\medskip\noindent{\bf Informal description of the classical algorithm.}
We will consider a priority function $p: S\to [d]$.
The objective $\Phi$ for player~1 is the weak-parity objective
$\wpe(p)$ and the objective for player~2
is the complementary objective $\Psi=\wpo(p)$.
The algorithm proceeds by computing attractors and 
removing the attractors from the game graph and proceeds on 
the subgame graph.
At iteration $i$, we denote the game graph by $G^i$ and 
the state space as $S^i$ and the set of edges of $G^i$ as
$E^i$.
At iteration $i$, the attractor set to the set of states 
of priority $i$ in $G^i$ (i.e., attractor to 
$p^{-1}(i) \cap S^i$) is computed.
If $i$ is even, the set is included in the winning set for 
player~1, and otherwise it is included in the winning 
set for player~2 and the set is removed from the game graph 
for the next iterations.

\begin{algorithm}[t]
\caption{\bf Classical algorithm for Weak-parity Objectives}
\label{algorithm:classical}
{ 
\begin{tabbing}
aa \= aa \= aaa \= aa \= aa \= aa \= aa \= aa \kill
\> {\bf Input :} A 2-player game graph $G=((S,E),(S_1,S_2))$ and 
 priority function $p:S \to [d]$. \\
\> {\bf Output:} A partition $(W_1,W_2)$ of  $S$. \\
\> 1. $G^0:=G$; $W_1:=W_2:=\emptyset$;\\
\> 2. {\bf for } ($i:=0; i<d; i:=i+1$) \\
\>\> 2.1. $A_i= \attr_{(i\mod 2 + 1)}(p^{-1}(i) \cap S^i, G^i)$; \\
\>\> 2.2  $W_{(i\mod 2 + 1)} = W_{(i \mod 2 +i)} \cup A_i$; \\
\>\> 2.3. $G^{i+1} =G^i \setminus A_i$; \\
\> {\bf end for} \\ 
\> 3. {\bf return } $(W_1,W_2)$;
\end{tabbing}
}
\end{algorithm}

\medskip\noindent{\bf Correctness.}
The following theorem states the correctness of Algorithm~\ref{algorithm:classical}. 

\begin{theorem}[Correctness]\label{thrm:classical}
Given a game graph $G=((S,E),(S_1,S_2))$ and priority function 
$p:S \to [d]$ we have 
	\[
	W_1=W_1(\wpe(p)); \qquad 
	S\setminus W_1=W_2= W_2(\wpo(p)),
	\]
	where $(W_1,W_2)$ is the output of Algorithm~\ref{algorithm:classical}. 
\end{theorem}
\begin{proof}
Observe that in the game graph $G^i$ we have $S^i \subseteq \bigcup_{j \geq i} 
p^{-1}(j)$, i.e., the priorities in $G^i$ are at least $i$.
Let us denote by $W_1^i$ and $W_2^i$ the sets $W_1$ and $W_2$ at the end 
of iteration $i-1$ of Algorithm~\ref{algorithm:classical}.
Then for all $s \in S^i \cap S_1$ we have $E(s) \subseteq S^i \cup W_2^i$
and for all $s \in S^i \cap S_2$ we have $E(s) \subseteq S^i \cup W_1^i$.
We prove by induction that the following two conditions hold
\[
W_1^i \subseteq W_1^G\big(\wpe(p) \cap \set{\pat \mid \min (p(\Occur(\pat))) < i }\big);
\]
\[
W_2^i \subseteq W_2^G\big(\wpo(p)) \cap \set{\pat \mid \min (p(\Occur(\pat))) < i }\big).
\]
The base case is trivial and we now prove the inductive case.
For $i$ even, for a state $s \in A_i$, the attractor strategy $\straa$ 
for player~1 in $G^i$ to reach $p^{-1}(i) \cap S^i$ and then choosing edges 
in $S^i$, ensures that for all strategies $\strab$ for player~2 we have 
\[
\pat(s,\straa,\strab) \in \big(\wpe(p) \cap \set{\pat \mid \min(p(\Occur(\pat))) \leq i } \big) 
\cup \Reach{W_1^i}.
\]
By the inductive hypothesis it follows that 
\[
A_i \subseteq W_1^G\big(\wpe(p) \cap \set{\pat \mid \min(p(\Occur(\pat))) < i+1}\big).
\]
Similarly, it follows for $i$ odd that  
$A_i \subseteq W_2^G\big(\wpo(p) \cap \set{\pat \mid \min(p(\Occur(\pat))) < i+1}\big)$.
The desired result follows.
\qed
\end{proof}

\medskip\noindent{\bf Running time analysis.}
In the running time analysis we will denote by $n$ the number of 
states, and by $m$ the number of edges in the game graph.
The naive analysis of the running time of Algorithm~\ref{algorithm:classical}
yields a $O( d \cdot m)$ running time analysis.
This is because the loop of step~2 runs for $d$ times, and each iteration
can be computed in $O(m)$ time.
Since $d$ can be $O(n)$, the worst case bound of the naive analysis is 
$O(n \cdot m)$, which is quadratic.
We will now present a linear-time analysis of the algorithm.
The two key issues in the running time analysis of the algorithm 
is to analyze the computation of the attractors (step 2.1 of the
algorithm) and obtaining the target sets 
$p^{-1}(i) \cap S^i$ in the attractor computation.
We now analyze the running time of the algorithm addressing the two above
issues.

\medskip\noindent{\bf The attractor computations.}
We first argue that the attractor computation over all iterations can be 
done in $O(m)$ time.
To prove this claim we observe that the sets $A_i$ computed at step 2.1
of the algorithm satisfies that 
$A_i \cap A_j =\emptyset$ for $i \neq j$, (since the set $A_i$ once 
computed is removed from the game graph for further iterations).
Let us consider the set $E_{A_i} = E^i \cap (S^i \times A_i)$ of edges.
Then for $i \neq j$ we have $E_{A_i} \cap E_{A_j} =\emptyset$.
By Lemma~\ref{lemm:attrcompute} it follows that the $i$-th iteration of
the attractor can be computed in $O(|E_{A_i}|)$ time.
Hence the total time for attractor computations over all 
iterations is 
\[
\sum_{i=0}^{d-1} O(|E_{A_i}|) =O(|E|)=O(m),
\] 
where the first equality follows since the edges $E_{A_i}$ and $E_{A_j}$ are
disjoint for $i \neq j$.

\medskip\noindent{\bf Obtaining the target sets.}
We will now argue that the target sets $p^{-1}(i) \cap S^i$ can be 
computed in $O(n)$ time over all iterations.
Without loss of generality we assume that the set of 
states $S$ are 
numbered $0,1,\ldots,n-1$ and the priority function 
$p:S \to [d]$ is given as an array 
$P[0 .. n-1]$ of integers such that $P[i]=p(i)$.
The procedure for obtaining the target sets will 
involve several steps. 
We present the steps below.

\begin{enumerate}
\item \emph{Renaming phase.} 
First, we construct a renaming of the states such that states in 
$p^{-1}(i)$ are numbered lower than $p^{-1}(j)$ for $i<j$.
Here is a $O(n)$ time procedure for renaming.

\begin{enumerate}
\item Consider an array of counters $\ct[0 .. d-1]$ all initialized to~$0$,
	and arrays $A[0],A[1],\ldots,A[d-1]$ (each $A[i]$ is an 
	array and will contain states of priority $i$).

\item The first step is as follows. 
\begin{quote}
{\bf for} ($i:=0; i<n; i:= i+1$) \\
\hspace*{1em} \{ \\
\hspace*{2em} $k=P[i]$; $j=\ct[k]$; \\
\hspace*{2em} $A[k][j]=i$; \\ 
\hspace*{2em} $\ct[k] =\ct[k]+1$; \\
\hspace*{1em} \} \\
\end{quote}
This step assigns to the array in $A[i]$ the set of states with 
priority $i$ (in the same relative order) and also works 
in $O(n)$ time. The counter $\ct[i]$ is the number of states with
priority $i$.

\item \emph{The renaming step.} We now construct arrays $B$ and $C$ 
in $O(n)$ time to store renaming and the inverse renaming. 
For simplicity let us assume $\ct[-1]=0$ and the procedure is as follows.
\begin{quote}
{\bf for} ($i:=0; i < d; i:=i+1$) \\
\hspace*{1em} {\bf for} ($j:=0; j < \ct[i]; j:=j+1$) \\
\hspace*{2em} \{ \\
\hspace*{3em} $B[\ct[i-1]+j]= A[i][j]$; \\ 
\hspace*{3em} $C[A[i][j]] =\ct[i-1]+j;$ \\
\hspace*{2em} \} \\
\end{quote}	
This creates the renaming such that for $B[0 .. \ct[0]-1]$ are
states of priority~0, then we have states of priority~1 for 
$B[\ct[0] .. \ct[1]-1]$, and so on.
The array $C$ stores the inverse of the renaming, i.e., 
if $B[i]=j$, then $C[j]=i$.
Moreover, though it is a nested loop, 
since $\sum_{i=1}^{d-1} \ct[i]=n$ this procedure also works
in $O(n)$ time.
\end{enumerate}

\item In the renaming phase we have obtained in $O(n)$ time a renaming in the 
array $B$ and the inverse renaming in the array $C$. 
Since renaming and its inverse, for a given state, 
can be obtained in constant time\footnote{We assume the random access model, 
and an element in the arrays $B$ and $C$ can be accessed in constant time.} 
we can move back and forth the renaming without increasing the time complexity 
other than in constants. 
We now obtain the set of states as targets required for the attractor 
computation of step 2.1 of Algorithm~\ref{algorithm:classical} 
in total $O(n)$ time across the whole computation. 
First, we create a copy of $B$ as an array $D$, and keep a global counter 
called $g$ initialized to~0.
We modify the attractor computation in step 2.1 such that in the attractor computation
when a state $j$ is removed from the game graph, then $D[k]$ 
is set to $-1$ such that $D[k]=j$,  (the entry of the array $D$ that 
represent state $j$ is set to $-1$).
This is simply done as follows $D[C[j]]=-1$. 
This is a constant work for a state and hence the extra work in the
attractor computation of step 2.1 across the whole computation is $O(n)$. 
The computation to obtain the target for priority~$i$ (i.e., $p^{-1}(i) \cap S^i$),
denoted as procedure \emph{ObtainTargets}, is described below.
The procedure \emph{ObtainTargets} is called by Algorithm~\ref{algorithm:classical}
with parameter $i$ in step 2.1 to obtain $p^{-1}(i) \cap S^i$.

\begin{enumerate}
\item We have the global counter $g:=0$ (initialized to $0$) and the value 
	of the global counter  persists across calls to the procedure 
	\emph{ObtainTargets}. 
	We present the pseudocode for the procedure \emph{ObtainTargets} to obtain 
	in an array $T$ the set $p^{-1}(i) \cap S^i$ of states.
	The procedure assumes that when \emph{ObtainTargets}$(i)$ is invoked
	we have $g=0$, if $i=0$, and for $i>0$, we have 
	$g=\sum_{j=0}^{i-1} \ct[j]$.
	Also, for all $j \in S \setminus S^i$ we have $D[C[j]]=-1$ 
	(the set of states in $S \setminus S^i$ is set to $-1$ in the attractor 
	computation). 	
	The procedure invoked with $i$ returns $T$ as an array with states in 
	$p^{-1}(i) \cap S^i$, and sets $g=\sum_{j=0}^{i} \ct[j]$.
	
\begin{quote}
\emph{ObtainTargets}($i$) \\
\hspace*{1em} $k:=0$; \\
\hspace*{1em} {\bf for} ($j:=0; j < \ct[i]-1; j:=j+1$) \\
\hspace*{2em}  \{ \\
\hspace*{3em} {\bf if} ($D[j+g] \neq -1$) \\
\hspace*{4em}  \{ $T[k] = D[j+g]$; $k:=k+1$; \} \\ 
\hspace*{2em} \} \\
\hspace*{1em} $g:= g + \ct[i]$; \\
\hspace*{1em} {\bf return} $T$. \\
\end{quote}
The work for a given $i$ is $O(\ct[i])$ and since 
$\sum_{i=0}^{d-1} \ct_{i}=n$, the total work to get the target 
sets over all iterations is $O(n)$.
\end{enumerate} 
\end{enumerate}
This completes the $O(n+m)=O(m)$ running time analysis for 
Algorithm~\ref{algorithm:classical}.
This yields the following result.

\begin{theorem}[Running time]\label{thrm:classical-run}
Given a game graph $G=((S,E),(S_1,S_2))$ and priority function 
$p:S \to [d]$, the sets
$W_1(\wpe(p))$ and $W_2(\wpo(p))$ can be computed in $O(m)$ time,
where $m=|E|$.
\end{theorem}

\subsection*{Acknowledgments.} I thank Florian Horn for useful comments.

\end{document}